\documentclass[aps,prb,twocolumn,superscriptaddress,showpacs]{revtex4}

\usepackage{graphicx}
\usepackage{dcolumn}
\usepackage{bm}

\begin{document}

\title{Paramagnetic ground state with field-induced partial order in
Nd$_3$Ga$_5$SiO$_{14}$ probed by low-temperature heat transport}

\author{Q. J. Li}
\affiliation{Hefei National Laboratory for Physical Sciences at
Microscale, University of Science and Technology of China, Hefei,
Anhui 230026, People's Republic of China}

\author{Z. Y. Zhao}
\affiliation{Hefei National Laboratory for Physical Sciences at
Microscale, University of Science and Technology of China, Hefei,
Anhui 230026, People's Republic of China}

\author{H. D. Zhou}
\affiliation{National High Magnetic Field Laboratory, Florida
State University, Tallahassee, Florida 32306-4005, USA}

\author{W. P. Ke}
\affiliation{Hefei National Laboratory for Physical Sciences at
Microscale, University of Science and Technology of China, Hefei,
Anhui 230026, People's Republic of China}

\author{X. M. Wang}
\affiliation{Hefei National Laboratory for Physical Sciences at
Microscale, University of Science and Technology of China, Hefei,
Anhui 230026, People's Republic of China}

\author{C. Fan}
\affiliation{Hefei National Laboratory for Physical Sciences at
Microscale, University of Science and Technology of China, Hefei,
Anhui 230026, People's Republic of China}

\author{X. G. Liu}
\affiliation{Hefei National Laboratory for Physical Sciences at
Microscale, University of Science and Technology of China, Hefei,
Anhui 230026, People's Republic of China}

\author{L. M. Chen}
\affiliation{Department of Physics, University of Science and
Technology of China, Hefei, Anhui 230026, People's Republic of
China}

\author{X. Zhao}
\affiliation{School of Physical Sciences, University of Science
and Technology of China, Hefei, Anhui 230026, People's Republic of
China}

\author{X. F. Sun}
\email{xfsun@ustc.edu.cn}

\affiliation{Hefei National Laboratory for Physical Sciences at
Microscale, University of Science and Technology of China, Hefei,
Anhui 230026, People's Republic of China}

\date{\today}

\begin{abstract}

We study the low-temperature heat transport of
Nd$_3$Ga$_5$SiO$_{14}$, which is a spin-liquid candidate, to probe
the nature of ground state and the effect of magnetic field on the
magnetic properties. The thermal conductivity ($\kappa$) shows a
purely phononic transport in zero field. The external magnetic
field along the $c$ axis induces a dip-like behavior of
$\kappa(H)$, which can be attributed to a simple paramagnetic
scattering on phonons. However, the magnetic field along the $ab$
plane induces another step-like decrease of $\kappa$. This kind of
$\kappa(H)$ behavior is discussed to be related to a field-induced
partial order, which yields low-energy magnetic excitations that
significantly scatter phonons. These results point to a
paramagnetic ground state that partial magnetic order can be
induced by magnetic field along the $ab$ plane, which is also
signified by the low-$T$ specific heat data.

\end{abstract}


\maketitle

\section{Introduction}

Geometrically frustrated magnetic materials with competing spin
interactions give rise to macroscopic degeneracy of ground state.
They have displayed abundant exotic magnetic phenomena and
introduced important concepts to condensed matter
physics.\cite{Levi, Review, Broholm, Shores, Gardner, Okamoto} The
recently discovered rare-earth Langasites, $R_3$Ga$_5$SiO$_{14}$
($R$ = Nd, Pr), with two-dimensional geometrical frustration have
attracted much attention due to their possible spin-liquid
state.\cite{Robert, Bordet, Zhou}

Nd$_3$Ga$_5$SiO$_{14}$ (NGSO) crystallizes in the trigonal space
group $P$321, and Nd$^{3+}$ spins ($S = 9/2$) occupy a network of
corner-sharing triangles, forming a distorted kagom\'{e}-like
lattice in the $ab$ plane stacked along the $c$ axis, which is
topologically equivalent to a perfect one when only the
nearest-neighboring interaction is considered.\cite{Bordet} The
magnetic anisotropy associated with crystal field effect was
confirmed from the magnetic susceptibility and nuclear magnetic
resonance (NMR) measurements, which displayed a distinct crossover
from the easy plane (kagom\'{e} plane) to the easy axis ($c$ axis)
below 33 K.\cite{Bordet, Zorko} The experimental investigations on
the ground state of NGSO seemed to be rather confusing. The muon
spin relaxation ($\mu$SR) and neutron scattering experiments did
not detect any phase transition or long-range magnetic order down
to milliKelvin temperatures in zero field.\cite{Robert, Zhou,
Zorko} In some earlier works, the magnetic susceptibility showed
that the Curie-Weiss temperature is about -52 K or -62 K and
therefore the frustration factor is large ($> 1300$), indicating a
spin-liquid ground state with rather strong spin
interactions.\cite{Bordet, Zhou} However, by measuring the
low-lying crystal-field levels of Nd$^{3+}$ via the
magneto-optical spectroscopy, it was found that the exchange
interaction of Nd$^{3+}$ spins is surprisingly small, leading to a
revised frustration factor of $\sim$ 1.\cite{Xu} This finding is
essentially consistent with the single-ion quantum relaxation at
low temperatures revealed by the neutron scattering.\cite{Simonet}
These results ruled out NGSO as a spin liquid at least down to 100
mK. So the ground state of NGSO remains to be fully understood. In
addition, the effect of magnetic field on the ground state of NGSO
has been probed by magnetization and neutron scattering
measurements.\cite{Bordet, Zhou, Simonet} At 1.6 K, the
magnetization for $H \parallel c$ and $H \perp c$ showed a simple
spin polarization, but the saturated moments are only half of the
free Nd$^{3+}$ moment, which is still an open
question.\cite{Bordet, Zorko, Simonet}

Low-temperature heat transport has recently been found to be an
effective way to probe the magnetic excitations and the
field-induced quantum phase transitions.\cite{Yamashita1,
Yamashita2, DTN, Sologubenko1} In this work, we study the heat
transport properties of NGSO at low temperatures down to 0.3 K and
in magnetic field up to 14 T. The thermal conductivity ($\kappa$)
shows a typical phononic heat transport behavior in zero field and
at very low temperatures, while it exhibits remarkably anisotropic
magnetic-field dependence. The magnetic field along the $c$ axis
causes a low-field dip in $\kappa(H)$ isotherms and the high-field
$\kappa$ are the same as the zero-field values, which can be
simply understood as a paramagnetic scattering on phonons. Whereas
the $\kappa(H)$ isotherms in magnetic field along the $ab$ plane
show sharp and step-like decreases, which indicates a
field-induced magnetic transition. These transport results point
to a paramagnetic ground state of NGSO but a field-induced partial
order for $H \parallel ab$, which is consistent with what the
low-$T$ specific heat data suggest.

\section{Experiments}

High-quality NGSO single crystals are grown by the floating-zone
technique in flowing argon and oxygen mixture with the ratio of
10:1.\cite{Bordet, Zhou} To reduce the bubble during the growth,
the sintered feed rod is firstly pre-melted with a rate of 25
mm/h. Then the single-crystal rods (5 mm in diameter and 9 mm in
length) are successfully obtained with a low growth rate of 1
mm/h. It should be noted that this fast-scan procedure is very
useful to improve the quality of NGSO crystals in the aspects of
both the size and the crystallinity.\cite{Bordet, Zhou} The
samples for thermal conductivity measurements are cut into
parallelepiped shape with a typical size of 3 $\times$ 0.6
$\times$ 0.15 mm$^3$ along either the $a$ axis or the $c$ axis.
The thermal conductivity is measured at low temperatures down to
0.3 K using a conventional steady-state technique, which has been
described in details elsewhere.\cite{DTN, HoMnO3, GdFeO3, Nd2CuO4}
The specific heat is measured by the relaxation method in the
temperature range from 0.4 to 30 K using a commercial physical
property measurement system (PPMS, Quantum Design).

\section{Results and discussion}

\begin{figure}
\includegraphics[clip,width=7.0cm]{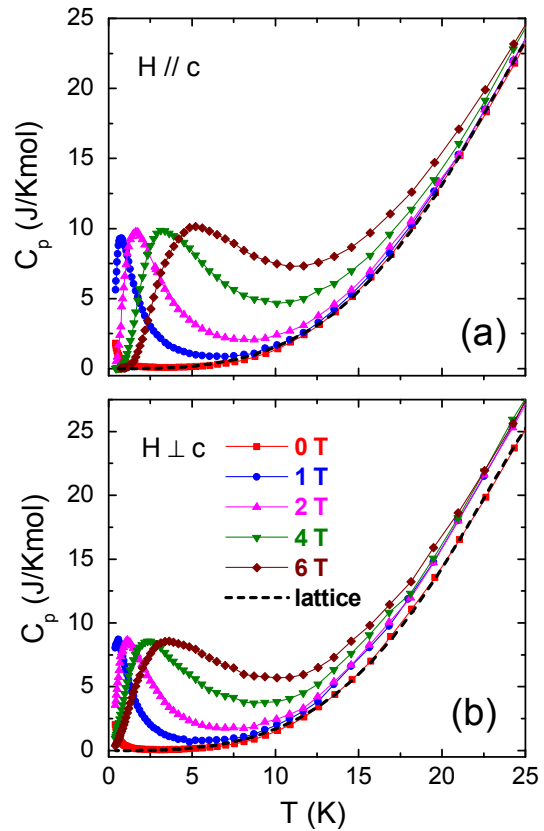}
\caption{(Color online) Temperature dependencies of the
low-temperature specific heat of NGSO single crystals in the
magnetic field applied along (a) and perpendicular (b) to the $c$
axis. The dashed black lines show the fitting of the high-$T$
zero-field data by using the Debye model of lattice specific
heat.}
\end{figure}

Before presenting the heat transport results of NGSO single
crystals, we study the low-$T$ specific heat in details. Figure 1
shows the temperature dependencies of the specific heat in zero
field and in magnetic fields for $H \parallel c$ and $H \perp c$.
The zero-field specific heat shows a upturn at low temperatures
below 2 K, which has been observed in previous work and is
apparently of magnetic origin. The distinct peaks develop in
various magnetic fields both for $H \parallel c$ and $H \perp c$
with the peak position shifting to higher temperatures with
increasing field. This is a characteristic behavior of the
Schottky anomaly, which commonly exists in many magnetic
materials. The phononic specific heat can be estimated by fitting
the high-$T$ data in zero field using the low-frequency expansion
of the Debye function $C_p = \beta T^3 + \beta_5T^5 +
\beta_7T^7$,\cite{Tari} as shown by the dashed lines in Fig. 1.
The best fitting gives $\beta = 1.5\times{10^{-3}}$ J/K$^4$mol,
$\beta_5 = 1.29\times{10^{-6}}$ J/K$^6$mol, and $\beta_7 =
-1.93\times {10^{-9}}$ J/K$^8$mol. The magnetic specific heat
$C_m$ can be obtained by subtracting the lattice contribution from
the total $C_p$, as shown in Fig. 2. The Schottky anomaly in NGSO
can be simply interpreted as the splitting of a ground-state
doublet of Nd$^{3+}$ with the effective spin $S$ = 1/2. Therefore,
the data in magnetic fields are tried to fitted by using the
two-level Schottky anomaly\cite{Tari, PLCO}

\begin{equation}\label{1}
C_{sch}=N(\frac{\Delta}{k_BT})^{2}\frac{e^{\Delta/k_BT}}{(1+e^{\Delta/k_BT})^{2}},
\label{SH}
\end{equation}
where $\Delta$ is the gap value for the Zeeman splitting of the
Nd$^{3+}$ Kramers doublet. The concentrations of free spins is
$N/R$, with $R$ the universal gas constant.

\begin{figure}
\includegraphics[clip,width=7.0cm]{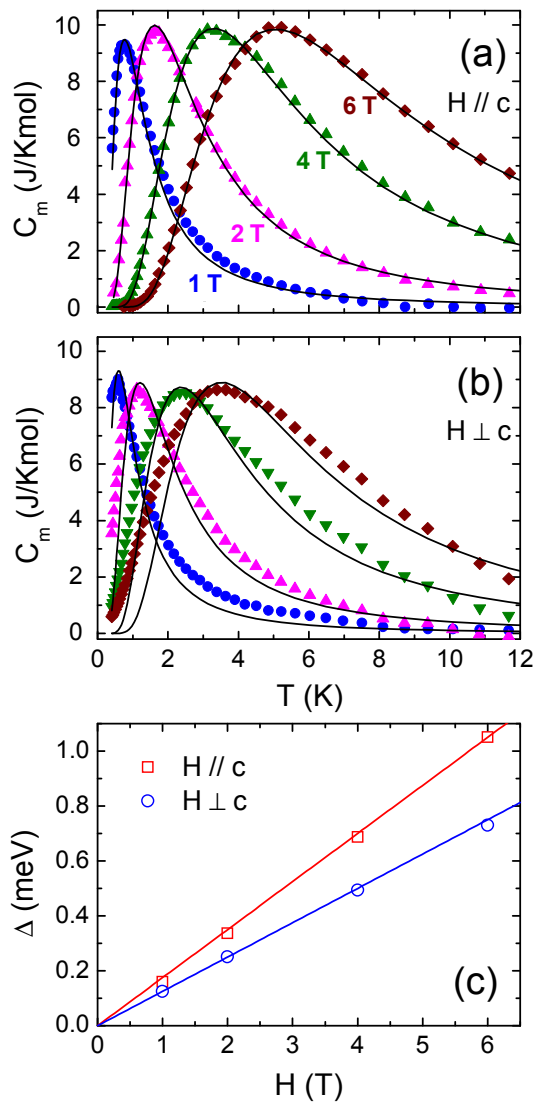}
\caption{(Color online) Temperature dependencies of magnetic
specific heat $C_m$ in field applied along (a) and perpendicular
to (b) the $c$ axis, respectively. The solid lines correspond to
the fitting results by a simple two-level Schottky model. (c) The
relationship between the fitting parameter $\Delta$ and magnetic
field. The thin lines show linear dependencies $\Delta = 0.175H$
and $\Delta = 0.125H$ for $H \parallel c$ and $H \perp c$,
respectively.}
\end{figure}

The best fitting results using Eq. (\ref{SH}) are also shown in
Fig. 2. It can be seen that the magnetic specific heat for $H
\parallel c$ can be fitted very well using the two-level Schottky
anomaly, indicating that the spins behave mainly like a free
system. The fitting parameter $\Delta \approx 0.175H$ for $H
\parallel c$, as shown in Fig. 2(c), is in good agreement
with previous results obtained from inelastic neutron scattering
spectrum.\cite{Zhou, Simonet} Another fitting parameter $N$ = 22.3
J/Kmol is nearly $3R$, which indicates that almost all the
Nd$^{3+}$ moments behave like paramagnetic spins in magnetic field
along the $c$ axis. For $H \perp c$, the similar data analysis are
also performed and shown in Fig. 2. The fitting parameter $\Delta
\approx 0.125H$ suggests the anisotropic Land\'e factor. The
parameter $N$ = 20.4 J/Kmol is rather consistent with the value
for $H \parallel c$. However, it is notable that there is a pretty
larger discrepancy between the fitting curves and the specific
heat data at very low temperatures, particularly for $H
>$ 1 T, as shown in Fig. 2(b). This means that at very low
temperatures, there must be some other kind of magnetic
contribution to the specific heat. These data suggest that the
ground state of NGSO is likely paramagnetic rather than
spin-liquid like. The unexpected phenomenon is that the magnetic
field along the $ab$ plane induces some magnetic excitations or
spin fluctuations, which indicate some kind of field-induced
partial or short-range magnetic order (no sign for long-range
order in the specific heat). As we can see below, such kind of
in-plane field induced order has remarkable impact on the low-$T$
heat transport.

\begin{figure}
\includegraphics[clip,width=8.0cm]{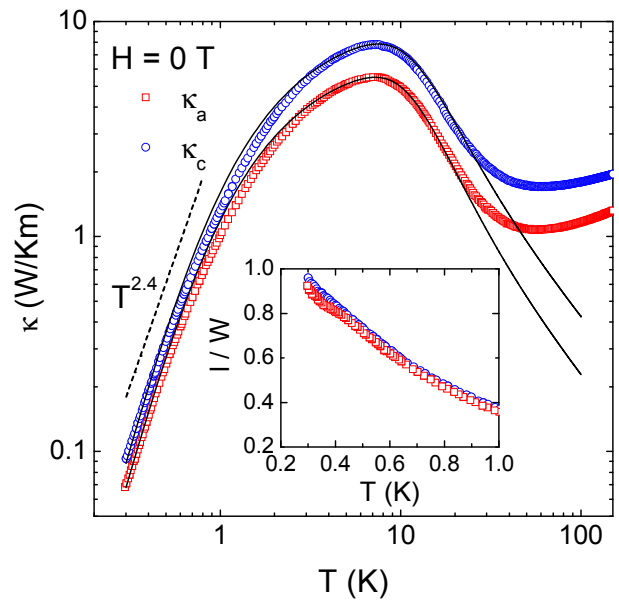}
\caption{(Color online) Temperature dependencies of thermal
conductivity along the $a$ axis and the $c$ axis of NGSO single
crystals in zero field, respectively. The solid lines are the
fitting results using the Debye model. The inset shows the
temperature dependencies of the ratio of the phonon mean free path
$l$ to the averaged sample width $W$.}
\end{figure}

Thermal conductivity as a bulk measurement technique is a
sensitive probe to all elementary excitations that carry heat or
scatter heat carriers.\cite{Berman} Figure 3 shows the temperature
dependencies of $\kappa$ with heat flowing along the $a$ and $c$
axes in zero field. Both curves show phonon peaks at about 7 K, a
typical feature of insulating crystals, and exhibit rather weak
anisotropy. At subKelvin temperatures, a rough $T^{2.4}$
dependence is observed, indicating that $\kappa(T)$ behaviors are
close to the boundary scattering limit.\cite{Berman} In order to
judge whether the phonons are free from microscopic scattering at
subKelvin temperatures, we make an estimation of the mean free
path $l$ of phonons at low temperatures.\cite{GdFeO3, Nd2CuO4} The
phononic thermal conductivity can be expressed by the kinetic
formula $\kappa_{ph} = \frac{1}{3}Cv_pl$, where $C = \beta T^3$ is
the phonon specific heat at low temperatures and $v_p$ is the
average velocity. With $\beta$ value obtained from the above
fitting of the specific-heat data, the average sound velocity
$v_p$ = 2410 m/s and the Debye temperature $\Theta_D$ = 310 K are
calculated by using the expression $\Theta_D = \frac{\hbar
v_p}{k_B}(\frac{6\pi^2Ns}{V})^{\frac{1}{3}}$ and $\beta =
\frac{12\pi^4}{5}\frac{Rs}{\Theta_D^3}$, where $V$ is the volume
of crystal, $R$ is the universal gas constant, $N$ is the number
of molecules per mole, and $s$ is the number atoms in each
molecule. The inset to Fig. 3 shows the temperature dependencies
of the ratio $l/W$ for $\kappa_a$ and $\kappa_c$, where $W$ is the
averaged sample width. It is found that both the $l/W$ ratios are
rather close to 1 with lowering temperature down to 0.3 K,
indicating that the phonon heat transport of NGSO indeed
approaches the boundary scattering limit at very low temperatures
and the microscopic scatterings on phonons are nearly wiped out.

For NGSO, the $\mu$SR and neutron spectroscopy revealed that the
magnetic fluctuations slow down below 10 K but keep persistent
down to milliKelvin temperatures in zero field.\cite{Zhou,
Simonet} However, the $\kappa(T)$ curves do not show any clear
anomalies around 10 K, which suggests that the magnetic
excitations do not couple with phonons. Therefore, it is possible
to try a more quantitative analysis on the zero-field data by
using a classical Debye model of phonon thermal
conductivity,\cite{Berman, Sologubenko2, Ba3Mn2O8}
\begin{equation}\label{(2)}
\kappa_{ph}=\frac{k_B}{2\pi^2 v_p}(\frac{k_B}{\hbar})^3
T^3\int_0^{\Theta_D/T} \frac{x^4e^x}{(e^x-1)^2} \tau(\omega,T)dx,
\label{kappa}
\end{equation}
in which $x = \hbar\omega/k_BT$ is dimensionless, $\omega$ is the
phonon frequency, and $\tau(\omega,T)$ is the mean lifetime of
phonon. The phonon relaxation is usually defined as
\begin{equation}\label{(3)}
    \tau^{-1}=v_p/L + A\omega^4 + BT\omega^3\exp(-\Theta_D/bT) +
    \tau_{res}^{-1},
\end{equation}
where the four terms represent the phonon scattering by the grain
boundary, the point defects, the phonon-phonon Umklapp scattering,
and the resonant scattering, respectively. Since the $\kappa(T)$
curves do not show clear signature of resonant scattering, the
last factor can be neglected. The parameters $v_p$ and $\Theta_D$
are obtained from the specific-heat data and other parameters $L$,
$A$, $B$ and $b$ are free ones. It is turned out that the
zero-field thermal conductivity below 25 K is well simulated as
the blue lines shown in Fig. 3. The best fitting parameters are $L
= 3.7 \times 10^{-4}$ m, $A = 8.15 \times 10^{-41}$ s$^3$, $B =
9.8 \times 10^{-29}$ K$^{-1}$s$^2$, $b = 8.7$, and $L = 4.8 \times
10^{-4}$ m, $A = 5.5 \times 10^{-41}$ s$^3$, $B = 5.0 \times
10^{-29}$ K$^{-1}$s$^2$, $b = 9.0$ for $\kappa_a$ and $\kappa_c$,
respectively. The parameters $L$ are found to be in good
consistent with the averaged sample widths of these two samples,
which are $3.70 \times 10^{-4}$ m ($\kappa_a$) and $3.95 \times
10^{-4}$ m ($\kappa_c$), respectively. Note that the failure of
fitting at $T >$ 25 K can be related to either the magnetic
fluctuations or the contributions of optical phonons.\cite{Hess}
However, these complexities at relatively high temperatures are
not involved in the following discussion about the magnetic-field
effect on low-$T$ heat transport.

\begin{figure}
\includegraphics[clip,width=7.0cm]{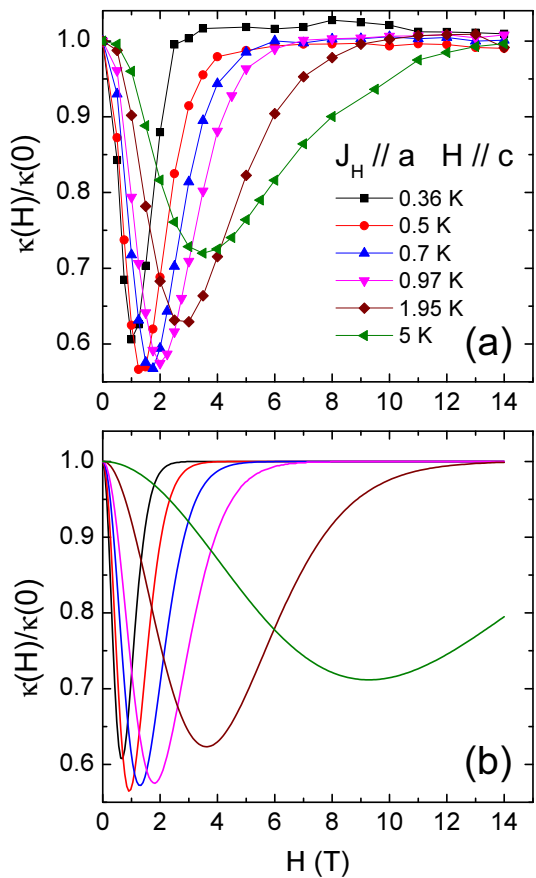}
\caption{(Color online) (a) Magnetic-field dependencies of thermal
conductivity of NGSO single crystal at low temperatures with
magnetic field along the $c$ axis. (b) The calculation results of
$\kappa(H)$ isotherms using the Debye model.}
\end{figure}

Detailed magnetic-field dependencies of the low-$T$ thermal
conductivity can provide useful information on the nature of the
ground state. Figure 4(a) shows the detailed magnetic field
dependencies of $\kappa_a$ for $H \parallel c$. The $\kappa(H)$
are strongly suppressed down to about 50\% at low magnetic field
and are recovered at high magnetic field, yielding a dip-like
feature. Note that the position of the minimum moves to high
magnetic field as increasing temperature. Based on the Schottky
contribution to the low-$T$ specific heat, the dip-like behaviors
of $\kappa(H)$ can be attributed to the phonon scattering by the
paramagnetic moments, that is, the field-induced splitting of
Nd$^{3+}$ Kramers doublet can cause a resonant scattering on
phonons, as we previously observed in Pr$_{1.3}$La$_{0.7}$CuO$_4$
and GdBaCo$_2$O$_{5+x}$.\cite{PLCO, GBCO} The Zeeman splitting of
the spin doublet is known from the above fitting of Schottky
anomaly for $H \parallel c$, which gives $\Delta(H) \approx
0.175H$. We can simulate the magnetic-field dependencies of
$\kappa(H)$ by using the Debye model for phonon thermal
conductivity,\cite{GBCO} as shown in Fig. 4(b). One can see that
most of $\kappa(H)$ isotherms are in good agreement with the
calculated ones except for some deviations for the high-$T$ data,
which might be due to the excitations of the higher-energy crystal
field levels. Therefore, the $c$-axis magnetic field seems to
affect the heat transport mainly by introducing some paramagnetic
scattering on phonons. This is in good correspondence with what
the specific-heat data suggest. It is worthy of pointing out that
the $\kappa_c(H)$ isotherms for $H \parallel c$ are the same as
those of $\kappa_a(H)$ shown in Fig. 4.

\begin{figure}
\includegraphics[clip,width=7.0cm]{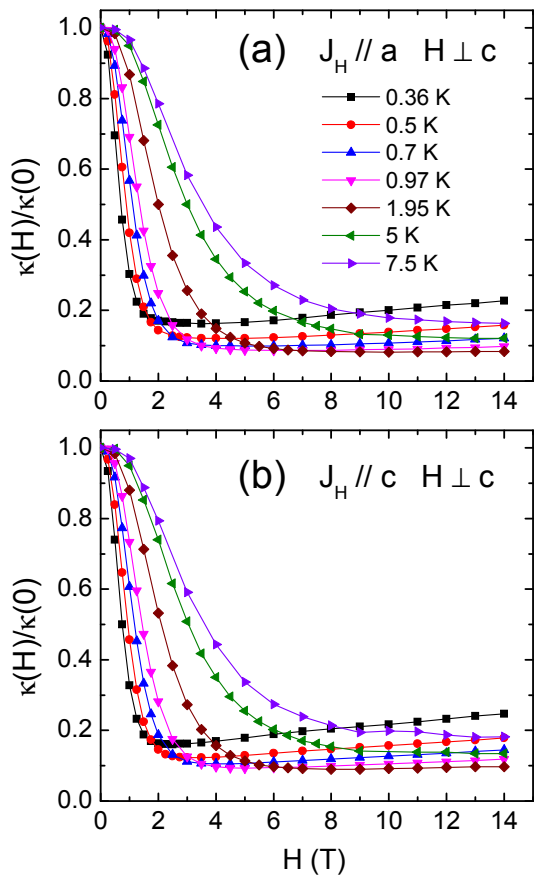}
\caption{(Color online) Magnetic-field dependencies of thermal
conductivities $\kappa_a$ (a) and $\kappa_c$ (b) of NGSO single
crystals at low temperatures with magnetic field perpendicular to
the $c$ axis.}
\end{figure}

The above paramagnetic scattering effect is known to be
qualitatively isotropic for magnetic field along different
directions.\cite{PLCO, GBCO} However, the heat transport behaves
very differently if the magnetic field is applied in the $ab$
plane. Figure 5 shows the magnetic-field dependencies of both
$\kappa_a$ and $\kappa_c$ for $ H \perp c$. It is easily found
that the $\kappa(H)$ behaviors are essentially the same for
different directions of heat current. All the $\kappa(H)$
isotherms show significant suppression down to 10--20\% of the
zero-field value. With lowering temperature, the decrease of
$\kappa$ becomes sharper at low fields. If a dip-like $\kappa(H)$
term from the paramagnetic scattering effect is also included, the
field dependence of Fig. 5 apparently has another contribution.
This can be clarified by a simple analysis. As some examples shown
in Fig. 6 (similar analysis can be done for other data), the
changes of $\kappa_a$ induced by paramagnetic scattering are
calculated in the same way as that for $H \parallel c$,
considering $\Delta(H) \approx 0.125H$. Then subtracting this part
of change from the raw data of $\kappa_a(H)$, the other
contribution to the field dependence can be roughly estimated. It
can be seen that at low temperatures, this additional term is
actually a step-like suppression, which suggests that the
$ab$-plane field induces some magnetic transition. This is
compatible with what the specific-heat data signify, as discussed
above. Although the quantitative description of the phonon
scattering by the magnetic excitations in such a broad temperature
region is not available now, the qualitative understanding can be
easily figured out. Apparently, the field-induced partial or
short-range order causes strong dynamic spin fluctuations that can
strongly scatter phonons.

\begin{figure}
\includegraphics[clip,width=7.0cm]{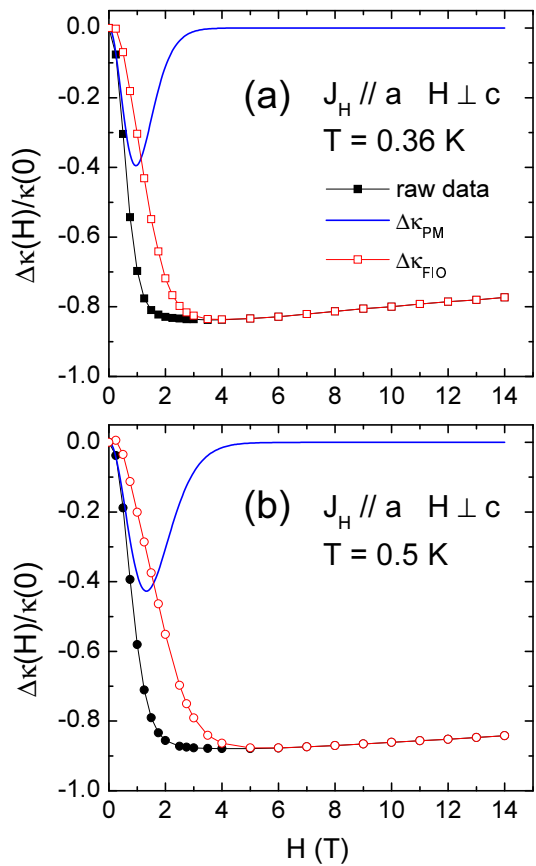}
\caption{(Color online) Magnetic-field-induced changes of
$\kappa_a$ at 0.36 K (a) and 0.5 K (b) for $H \perp c$. $\Delta
\kappa (H)$ is defined as $\kappa(H) - \kappa(0)$. The solid
symbols indicate the experimental data. The solid blue lines
represent the calculated changes of $\kappa$ by the paramagnetic
scattering, denoted as $\Delta \kappa_{PM}$. The open symbols show
the results of subtracting the paramagnetic scattering term from
the raw data, denoted as $\Delta \kappa_{FIO}$, which is discussed
to be related to some kind of field-induced magnetic order.}
\end{figure}

It should be pointed out that the in-plane field induced partial
order, suggested by the specific heat and thermal conductivity
data in the present work, however, has not been evidenced in the
earlier works. It is known that the low-$T$ magnetization showed a
simple spin polarization above $\sim$ 2 T for both $H
\parallel c$ and $H \perp c$,\cite{Bordet, Simonet} and the neutron
scattering also confirmed this phenomenon for $H \parallel
c$.\cite{Zhou} But the moments of the high-field $M(H)$ plateau
are only half of the free Nd$^{3+}$ moment, which is an open
question for the existing theories on the kagom\'e
antiferromagnet.\cite{Bordet, Zorko, Simonet} There have been no
other experimental investigation about the details for $H \perp
c$. Here, we propose one possibility that when the Nd$^{3+}$
moments are gradually aligned along the in-plane field, the
$c$-axis component of Nd$^{3+}$ moments may form some kind of AF
ordering. They could be anti-parallel either for the neighboring
sites or for the neighboring layers. This picture is not
inconsistent with the experimental results that the high-field
$M(H)$ plateau is lower for $H \perp c$ than that for $H \parallel
c$.\cite{Bordet, Simonet} Nevertheless, the validity of this
canted AF order and the underlying mechanism, which should be
related to the spin anisotropy, the crystal field effect, and the
spin frustration, calls for a deeper investigation.

\section{CONCLUSION}

In summary, the temperature and magnetic field dependencies of the
thermal conductivity and specific heat are studied for NGSO single
crystals. The zero-field thermal conductivity shows a purely
phononic heat transport behavior at low temperatures, without
showing the strong coupling between phonons and magnetic
excitations. A dip-like field dependence of $\kappa$ for magnetic
field along the $c$ axis can be explained by a simple paramagnetic
scattering on phonons; whereas a step-like field dependence of
$\kappa(H)$ for $H \perp c$ indicates a field-induced magnetic
transition at low temperatures. These behaviors are consistent
with the field dependencies of low-$T$ specific heat. The present
results demonstrate a paramagnetic ground state rather than the
spin liquid. However, the partial magnetic order induced by the
field along the $ab$ plane seems to be quite unusual and remains
to be further investigated both experimentally and theoretically.

\begin{acknowledgements}

This work was supported by the Chinese Academy of Sciences, the
National Natural Science Foundation of China, the National Basic
Research Program of China (Grant Nos. 2009CB929502 and
2011CBA00111), and the Fundamental Research Funds for the Central
Universities (Program No. WK2340000035).

\end{acknowledgements}

\end{document}